\pdfoutput=1
\documentclass[11pt]{article}

\usepackage{acl}

\usepackage{times}
\usepackage{latexsym}

\usepackage[T1]{fontenc}
\usepackage[utf8]{inputenc}

\usepackage{microtype}

\usepackage{inconsolata}

\usepackage{graphicx}
\usepackage{subfigure}

\usepackage{multirow}
\usepackage{listings}
\usepackage{xspace}
\usepackage{booktabs}
\usepackage{tikz}
\usepackage{pgfplots}
\usepgfplotslibrary{groupplots}

\usepackage{makecell}

\usepackage{amsmath}
\usepackage{amssymb}
\usepackage{mathtools}
\usepackage{amsthm}
\usepackage{algpseudocode}
\usepackage{algorithm}
\usepackage{xcolor} % 载入xcolor包
\usepackage{amsthm}
\theoremstyle{definition}

\usepackage{epsfig,amsmath,amsfonts,epsfig,multirow,makecell,caption,soul,csquotes,color,wrapfig,mathtools,bm,spverbatim,booktabs,tcolorbox,diagbox,todonotes}
\usepackage[e]{esvect}

\usepackage{caption}
\newenvironment{changemargin}[2]{\begin{list}{}{
	\setlength{\topsep}{0pt}\setlength{\leftmargin}{-8pt}
	\setlength{\rightmargin}{0pt}
	\setlength{\listparindent}{\parindent}
	\setlength{\itemindent}{\parindent}
	\setlength{\parsep}{0pt plus 1pt}
	\addtolength{\leftmargin}{#1}\addtolength{\rightmargin}{#2}
	}\item}
	{\end{list}}

\usepackage[first=0,last=9]{lcg}
\usepackage{colortbl}
\definecolor{Gray}{gray}{0.8}
\colorlet{Red}{red!10!white}
\colorlet{Blue}{blue!10!white}

\usepackage{hyperref}

\usepackage{scalerel}[2016/12/29]

\makeatletter
\providecommand{\leadsfrom}{%
  \mathrel{\mathpalette\reflect@squig\relax}%
}
\newcommand{\reflect@squig}[2]{%
  \reflectbox{$\m@th#1\leadsto$}%
}
\makeatother

\usepackage{amsmath,amsfonts,bm}

\def\eqref#1{equation~\ref{#1}}
\def\1{\bm{1}}

\DeclareMathAlphabet{\mathsfit}{\encodingdefault}{\sfdefault}{m}{sl}
\SetMathAlphabet{\mathsfit}{bold}{\encodingdefault}{\sfdefault}{bx}{n}

\definecolor{mygreen}{RGB}{114, 176, 99}
\definecolor{myorange}{RGB}{226, 145, 53}
\definecolor{myred}{RGB}{251, 231, 230}

\newcommand{\model}{{SecCoder}\xspace}

\title{SecCoder: Towards Generalizable and Robust Secure Code Generation}

\author{
    ~~Boyu Zhang$^{1}$
   ~~Tianyu Du$^{1}$\footnotemark[1]
   ~~Junkai Tong$^{2}$
   ~~Xuhong Zhang$^{1}$
   ~~Kingsum Chow$^{1}$\footnotemark[1]
   \AND
   ~~Sheng Cheng$^{1}$
   ~~Xun Wang$^{3}$
   ~~Jianwei Yin$^{1}$\\ 
   $^{1}$Zhejiang University \\
   $^{2}$Zhejiang University of Technology  \\
   $^{3}$Northeast Forest University \\
   \texttt{{\{zjuzby, zjradty\}@zju.edu.cn, 202103150317@zjut.edu.cn}}\\
   \texttt{{\{zhangxuhong, kingsum.chow\}@zju.edu.cn, taylorchang2016@gmail.com}}\\
\texttt{{deffitywang@outlook.com, zjuyjw@cs.zju.edu.cn}}\\
}

\tcbuselibrary{listingsutf8}

\begin{document}
\maketitle

\renewcommand{\thefootnote}{\fnsymbol{footnote}}
\footnotetext[1]{Corresponding author.}
% \footnotetext[2]{Second corresponding author.}
\renewcommand{\thefootnote}{\arabic{footnote}}

\lstdefinestyle{mystyle}{
    breaklines=true,
    basicstyle=\scriptsize,
    language={},
    frame={},
    escapeinside={{(*}{*)}},
}

\newtcblisting{mylisting}[2][]{
    arc=0pt, outer arc=0pt,
    title=#2, 
    colback=gray!5!white,
    colframe=black!75!black,
    fonttitle=\bfseries,
    listing only, 
    listing options={style=mystyle},
    % breakable,
}

\begin{abstract}
After large models (LMs) have gained widespread acceptance in code-related tasks, their superior generative capacity has greatly promoted the application of the code LM. Nevertheless, the security of the generated code has raised attention to its potential damage. Existing secure code generation methods have limited generalizability to unseen test cases and poor robustness against the attacked model, leading to safety failures in code generation. In this paper, we propose a generalizable and robust secure code generation method \model by using in-context learning (ICL) and the safe demonstration. The dense retriever is also used to select the most helpful demonstration to maximize the improvement of the generated code's security. Experimental results show the superior generalizability of the proposed model \model compared to the current secure code generation method, achieving a significant security improvement of an average of $7.20\%$ on unseen test cases. The results also show the better robustness of \model compared to the current attacked code LM, achieving a significant security improvement of an average of $7.74\%$. Our analysis indicates that \model enhances the security of LMs in generating code, and it is more generalizable and robust.
\end{abstract}

\section{Introduction}

\begin{figure}[t]
  \includegraphics[width=\columnwidth]{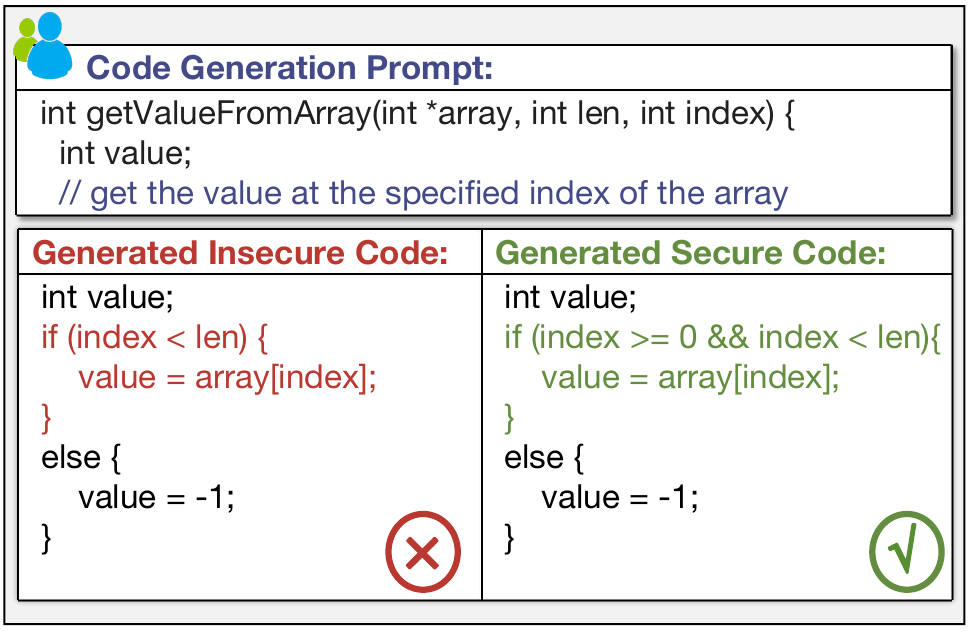}
  \caption{An illustration of secure code generation.}
  \label{fig:illustration}
\end{figure}

After large models (LMs) \cite{radford2019language, vaswani2017attention} achieved significant success, it has promoted the development of many code-related works such as code summarization \cite{parvez2021retrieval, ahmed2022few}, code repair \cite{xia2023automated, pearce2023examining}, code generation \cite{nijkamp2022codegen, wang2021codet5}. Nevertheless, the widespread use of LMs in code-related tasks has raised significant safety concerns. \citet{HammondPearce2023large} investigated the security of the code generated by GitHub Copilot \cite{githubcopilot} and found that about 40\% are vulnerable. \citet{siddiq2022securityeval} presented a manually curated dataset for code security evaluation. About 90\% of the code snippets generated by the LMs are vulnerable when manual inspection is used to check for security. The vulnerability poses a significant obstacle to code LMs' application in security-sensitive domains. To mitigate the vulnerabilities, the method of secure code generation has attracted increasing attention. Figure~\ref{fig:illustration} illustrates the secure code generation from Common Weakness Enumeration (CWE) \cite{cwe} serves as a broadly accepted category system for vulnerabilities.\footnote{\url{https://cwe.mitre.org/data/definitions/125.html}\label{fn:cwe-125}}

Thus far, extensive research has been conducted on enhancing the security of LMs \cite{ji2024beavertails, achiam2023gpt, qi2023fine}. Given the differences in security policies between the natural language processing (NLP) and the code, some safe alignment methods are specifically designed for code LMs \cite{JingxuanHe2023large}. Unfortunately, two crucial features of the secure code generation method have been ignored, which would severely compromise safety in practical applications.

The first is the generalizability to the unseen test cases. \citet{qi2023fine} proved that simply fine-tuning can inadvertently degrade the safety of LMs even without malicious intent. \citet{wei2024jailbroken} proposed that mismatched generalization is one of the critical failure modes of safety alignment. It occurs when safety training does not generalize to a domain for which capabilities exist. Compared to NLP, mismatched generalization is more prevalent in code generation since there are many kinds of vulnerabilities in code. For instance, the CWE \cite{cwe} has over 600 categories of vulnerabilities. The limited number of vulnerabilities in the secure code generation training dataset may lead to mismatched generalization in application \cite{JingxuanHe2023large}. Therefore, the lack of generalizability could cause safety failures, which limits the application of the secure code generation method.

The second is the robustness against the attacked model. There are many well-designed attacks on LMs \cite{schuster2021you, perez2022red, JingxuanHe2023large}. 
The experiments in \citet{JingxuanHe2023large} showed that their attack could not be easily defended using simple prompt perturbations without external knowledge. This indicates that the basic input preprocessing defenses are not robust against the attack. Therefore, a more powerful secure code generation method that is robust against such attacks is needed to generate more secure code in real-world applications.

To address the above challenges, in this work, we propose \model, a generalizable and robust secure code generation approach. Specifically, \model guides LMs to adapt swiftly to unseen test cases with the demonstration by leveraging the power of in-context learning (ICL) \cite{dong2022survey, min2021metaicl, iyer2022opt, wei2021finetuned, gu2023pre} ability. Additionally, \model enhances the robustness of secure code generation by providing an extra security codebase separately from the attacked model to guarantee the safe of the demonstration. \model retrieves the most helpful safe demonstration by using the retrieval capacity of the LMs to maximize \model's effectiveness. 

We employ several kinds of code LMs on a broad range of common vulnerabilities in the CWE \cite{cwe} to validate \model's generalizability and robustness. 
First, when evaluating the proposed model \model on the unseen test cases, the $12.07\%$ average increase in the security reveals \model's generalizability. Second, \model is more secure on unseen test cases than the state-of-the-art secure code generation method $\rm SVEN_{sec}$ \cite{JingxuanHe2023large} and the improvement of the security is $7.20\%$ on average, which reveals the generalizability of \model is better than the existing method.
Last, the security of the attacked code LM is increased by $7.74\%$ on average by using \model, which reveals the robustness of \model.
These results clearly demonstrate the power of \model. 

We also verify the functional correctness of \model since it is supposed to preserve the original LM's usefulness. We found that the functional correctness of \model is almost the same as the original LM despite not adopting any specific mechanisms to preserve the utility. It is a clear contrast to the existing method \cite{JingxuanHe2023large}, which carefully designed the mechanism to preserve the utility and paid a heavy price for the trade-off between the utility and the security. Our finding could inspire other researchers to find a more efficient and straightforward mechanism to preserve the utility of the LM during security hardening.

\textbf{Our Contributions.} Our main contributions can be summarized as follows:
\begin{itemize}
    \item We identify the primary limitations of the application of secure code generation methods: the generalizability to unseen test cases and the robustness against the attacked model.
    \item We propose \model that is a generalizable and robust secure generation method, which could preserve the utility without additional efforts and resources.
    \item Experiments show the effectiveness of \model in enhancing the generalizability and robustness of secure code generation. \model's generalizability outperforms the existing secure code generation method, and \model is robust against the existing attacked code LM.
\end{itemize}
\section{Related Work} 

\textbf{Security Risks of Code LMs.} Recent advances in pre-training technologies have facilitated the emergence of large-scale, pre-trained language models specifically tailored for code-related tasks, such as CodeX \cite{chen2021evaluating}, codeT5 \cite{wang2021codet5}, CodeGen \cite{nijkamp2022codegen}. Because the training dataset collected from open-source repositories like GitHub may include insecure code, the security of the code generated by LMs has raised serious concerns. \citet{HammondPearce2023large} evaluated the security in GitHub Copilot and found that roughly $40\%$ of the codes generated by it are insecure.
Inspired by this, \citet{JingxuanHe2023large} proposed SVEN to control the security of the generated code according to a binary property. Nevertheless, the security improvement reduces by $25\%$ when evaluating CodeGen-2.7B on the unseen test case, which indicates that the generalizability of SVEN is limited. The effectiveness of SVEN also implies that the existing LMs are fragile in code security because they could generate more vulnerabilities by using $\rm SVEN_{vul}$.

\textbf{In-Context Learning.} As model sizes and corpus sizes have expanded \cite{chowdhery2023palm, brown2020language, devlin2018bert}, LMs have exhibited the powerful ICL ability, the capability to learn a new task from a handful of contextual examples. Extensive research has demonstrated that LMs can accomplish many complicated tasks via ICL \cite{wei2022chain}. In contrast to supervised training, ICL represents a training-free learning paradigm. This approach significantly decreases computational expenses associated with adjusting the model to novel tasks. Therefore, ICL is beneficial for the generalizability.

\textbf{Retriever.} The retriever has attracted significant concerns recently \cite{guu2020retrieval, karpukhin2020dense, izacard2023atlas, borgeaud2022improving, asai2023retrieval} since it could assist people to retrieve the desired item automatically. There are two kinds of retrievers. One is the sparse retriever, such as BM25 \cite{robertson2009probabilistic}, which uses lexical matching, and the other is the dense retriever, which uses semantic matching. With the development of pre-trained models, there are increasingly off-the-shelf dense retrievers, such as INSTRUCTOR \cite{su2022one}. INSTRUCTOR is fine-tuned to efficiently adapt to diverse downstream tasks without additional training by jointly embedding the inputs and the task.
Several code-related tasks adopt retriever such as code autocompletion \cite{hashimoto2018retrieve}, code summarization \cite{parvez2021retrieval}, and code generation \cite{parvez2021retrieval}. Nevertheless, there is no widely agreed criterion for selecting a perfect demonstration. The existing research on retrieval strategies for secure code generation is still limited.

\begin{figure*}[t]
  \centering
  \includegraphics[width=1\textwidth]{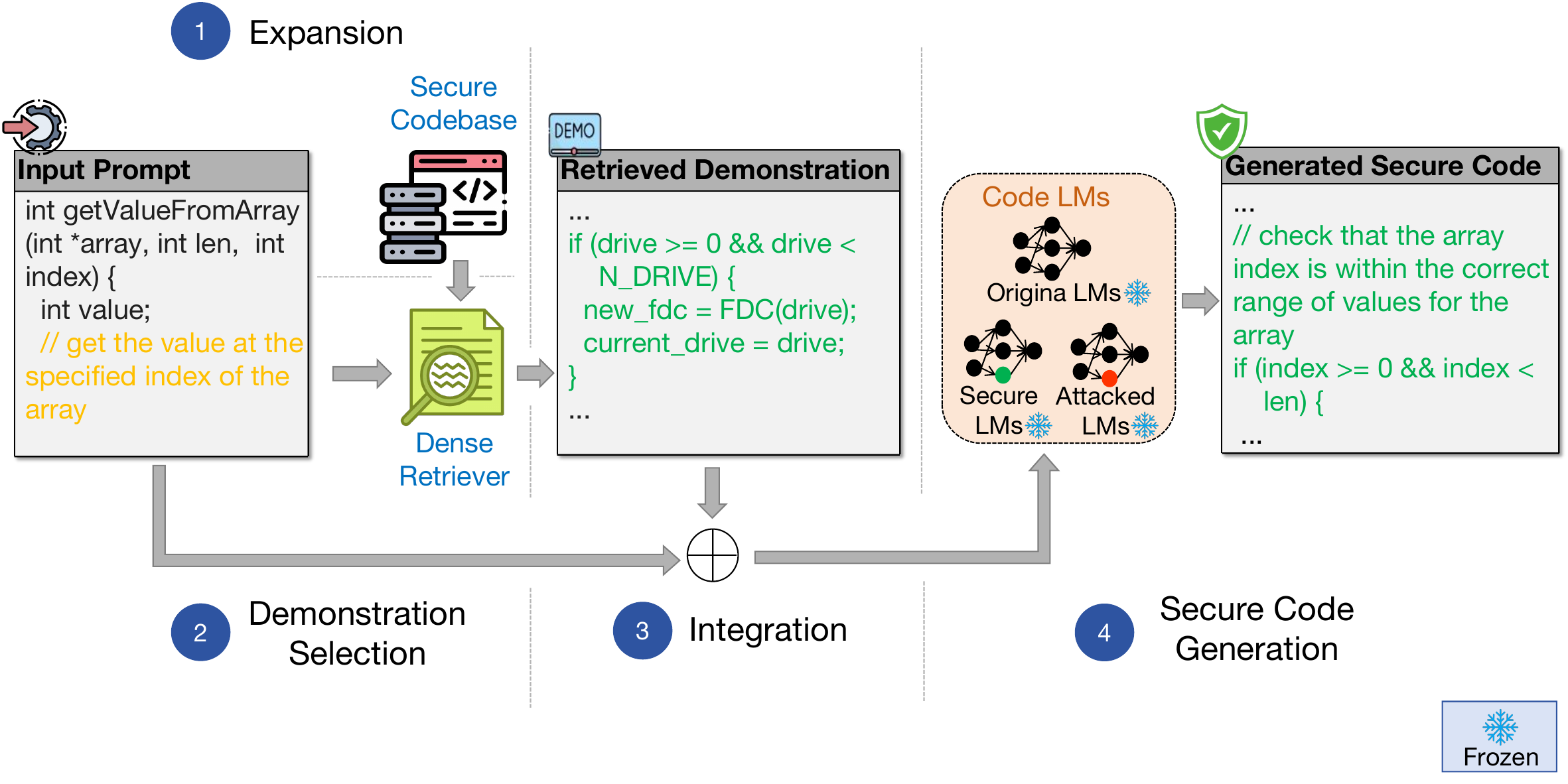}
  \caption{The framework of \model.}
  \label{fig:securityICL_framework}
\end{figure*}

\vspace{-10pt}
\section{Methodology}
\subsection{Overview}
In this section, we describe the proposed method in detail. As Figure~\ref{fig:securityICL_framework} depicts, we introduce \model, a novel method to enhance the generalizability and the robustness of the secure code generation method. It consists of four stages, each involving a different role of enhanced code security. Leveraging the LM's capabilities, \model is more generalizable and robust than the prior work.

\subsection{Problem Formulation}

Our ultimate goal is to generate a more secure code $y$ via:
\begin{equation}
y = \mathop{\arg\max}\limits_{y_k} LM({y_k|x}), 
\end{equation}
where $x$ is one of the prompts used to guide LMs to generate desired codes, consisting of an incomplete program and a functional description. $y_k$ indicates all possible results of $y$. Our approach is to optimize the process based on the following steps.

\subsection{Step 1: Expansion}
First, in order to improve the robustness, when a new vulnerability is found, fix and add it to the secure code database $S$ which contains a large collection of previous secure codes $\{s_1, s_2, \cdots, s_j, \cdots, s_{m}\}$, where $s_j$ denotes the $j$-th previous secure code and $m$ is the number of secure codes. The secure code database would be expanded to $S = \{s_1, s_2, \cdots, s_j, \cdots, s_{m}, s_{m+1}\}$.
The codes in the codebase are all secure to guarantee the security of the retrieved demonstration, which could improve the robustness of the proposed \model. The secure code could be collected from open-source platforms like GitHub or local projects. The latter method may be safer and more practical because it could resist malicious code on the open-source platform and avoid out-of-distribution problems.

\subsection{Step 2: Demonstration Selection}
Second, relying on the retrieval capability of the LM, we use the pre-trained embedding LM as the retriever to select the most helpful demonstration.
Given a prompt $x$, a dense retriever fetches the most relevant secure code $s_j$ in the codebase $S$ according to the relevance scoring function $f_\phi(x, s_j)$ parameterized by $\phi$. 
Specifically, the dense retriever encodes the prompt and the codes in the secure codebase into continuous vectors. Next, calculate their similarities and select the secure code that has the maximum similarity with the prompt. We choose cosine similarity since the critical character of the semantic is the direction of the vector instead of the length. Therefore, cosine distance is perfect for measuring the distance of embeddings.

\subsection{Step 3: Integration}
Third, leveraging the in-context learning capability of LMs improves the generalizability of \model. We show a demonstration to the LM and encourage the LM to generate more secure codes. The original input prompt $x$ is augmented with the retrieved secure code $s_j$ to form a new input prompt $\hat{x} = x \oplus s_j$, where $\oplus$ denotes the concatenation operation. The integration template is presented in Appendix \ref{experimental_detail}. The new input prompt would be sent to the code LMs.

\subsection{Step 4: Secure Code Generation}
Last, the new input prompt $\hat{x}$ would be used to generate the more secure code using the code LM.
We model the output of the code LM as a sequence of tokens \emph{i.e.}, $y$, which is supposed to be the more secure code that is generated according to the input $\hat{x}$:
\begin{equation}
y = \mathop{\arg\max}\limits_{y_k} LM({y_k|\hat{x}}),
\end{equation}
Algorithm \ref{alg:SecCoder} shows the complete algorithm for \model.

\vspace{3pt}
\begin{algorithm}[!t]
% \footnotesize
\caption{\model}\label{alg:SecCoder}
\textbf{Input:} $X = \{x_{i}\}_{i=1}^{n}$: secure code generation evaluation dataset; $S = \{s_{i}\}_{i=1}^{m}$: secure code demonstration dataset; $s_{m+1}$: new secure code which is fixed the vulnerability; $\mathsf{LM}$: code LM; $\mathsf{DenseRetriver}$: dense retriever; $\mathsf{cos\_sim}$: similarity calculation function\\
\textbf{Output:} $Y = \{y_{i}\}_{i=1}^{n}$: generated codes

\begin{algorithmic}[1]
\State $S\gets\{s_1, s_2, \cdots, s_j, \cdots, s_{m}, s_{m+1}\}$;
\For {$x \in X$}
\State $x_{emb}\gets \mathsf{DenseRetriver}(x)$;
\State $max_{sim} \gets 0$;
\For {$s \in S$}
\State $s_{emb}\gets \mathsf{DenseRetriver}(s)$;
\State $sim\gets \mathsf{cos\_sim}(x_{emb}, s_{emb})$;
\If {$sim > max_{sim}$}
\State $max_{sim} = sim$
\State $s_j \gets s$
\EndIf
\EndFor
\State $\hat{x} = x \oplus s_j$
\State $y = \mathop{\arg\max}\limits_{y_k} \mathsf{LM}({y_k|\hat{x}})$
\EndFor
\State \textbf{return} $Y = \{y\}$.
\end{algorithmic}
\end{algorithm}

\section{Experiments}

\subsection{Experimental Setup}

\subsubsection{Dataset}
Three kinds of datasets are required in the experiments: the training dataset used to train the baseline methods, the demonstration dataset consisting of secure codes used by \model, and the evaluation dataset used to evaluate the security of various secure code generation methods. 
% The evaluation datasets used in the three kinds of code LMs are the same. 
% The training datasets and the demonstration datasets are different. Therefore, we introduce the evaluation dataset first and then introduce the training and demonstration datasets separately.

\textbf{Training Dataset.} There are two training datasets required when training the baseline methods. One is used to train the state-of-the-art secure code generation method, and the other is used to train the state-of-the-art attacked code LM. The first dataset is constructed from \citet{fan2020ac}, and each data is labeled with a CWE tag. We use the dataset in \citet{fan2020ac} as the base dataset and remove the data whose CWE tag is the same as the data in the evaluation dataset to observe the generalizability of \model. Then, following our baseline $\rm SVEN_{sec}$ \cite{JingxuanHe2023large}, we randomly select 723 pairs of data from the rest.  
Second, we directly adopt the training dataset in \citet{JingxuanHe2023large} when training the attacked code LM to observe the robustness of the proposed method \model.

\textbf{Demonstration Dataset.}  
We construct two demonstration datasets from the existing datasets collected from GitHub to cover a broader range of vulnerabilities and real-world projects.
Each program in the two demonstration datasets is a function written in C/C++ or Python and related to a CWE that existed in the evaluation dataset. The first is constructed from the training dataset in \citet{JingxuanHe2023large} and used to observe the generalizability of \model. The second is constructed from the validation dataset in \citet{JingxuanHe2023large}, which is used to evaluate the robustness of \model on the attacked LM. The training dataset of the attacked LM and the evaluation dataset have the same CWE tags, but they have different secure codes. It simulates the situation in that the user is unaware of which data are used to attack the model. Deleting the secure programs according to the max context length, we get 596 secure codes in the first demonstration dataset and 63 secure codes in the second.

\textbf{Evaluation Dataset.} To evaluate \model, we use the evaluation dataset from \citet{JingxuanHe2023large}. Each evaluation data consists of an incomplete code snippet and a functional description. It has a CWE tag to identify the type of vulnerability that is prone to be produced when generating the code according to this evaluation data. The evaluation dataset covers 9 CWEs. This evaluation dataset is also used in \citet{HammondPearce2023large} and \citet{siddiq2022securityeval}, which proved that automatically measuring their security by using CodeQL \cite{CodeQLGitHub} is possible. 

\subsubsection{Models}
There are two kinds of models in \model, i.e., the code LM and the retriever.

\textbf{Code LMs.} 
% We study three security degrees of code LMs: the original LMs, the secure LMs, and the attacked LMs. 
We use CodeGen \cite{nijkamp2022codegen} with different sizes (350M, 2.7B, 6.1B), multi-head attention version SantaCoder (1.3B) \cite{allal2023santacoder}, and InCoder (6.7B) \cite{fried2022incoder}. 
% In the following parts, the original code LMs with None method indicate the above code LMs don't use any secure code generation method.
% Both Incode and SantaCoder were trained with different objectives.

\textbf{Retrievers.} The dense retriever used in \model is INSTRUCTOR \cite{su2022one}. We use INSTRUCTOR of two sizes in the experiments. Therefore, the suffix is used to distinguish the version of INSTRUCTOR. We use INSTRUCTOR-xl in \model-xl and INSTRUCTOR-large in \model-large. 
% The detail would be provided in Section~\ref{sec:impact_of_parameter_number_of_retriever}.

\subsubsection{Baselines}
The "None" method refers to the original code LM that does not employ any security mechanisms. To validate the generalizability of \model, we compare it with the state-of-the-art method $\rm SVEN_{sec}$ \cite{JingxuanHe2023large}.
To validate the robustness of \model, the adversarial testing method $\rm SVEN_{vul}$ \cite{JingxuanHe2023large} is used to attack the code LMs to reduce the security of the original LMs. Then, we observe whether the proposed method \model could be robust against the attacked model. 
% The attacked LMs with None method indicate they don't use any secure code generation method.
In the ablation study, we also compare \model with different retrieval strategies, such as random strategy and sparse retriever. BM25 \cite{robertson2009probabilistic} is selected as the sparse retriever.

\subsubsection{Metrics}

\textbf{Security Evaluation.}
We sample 25 completions and filter out the duplicates or the codes that have errors while compiling or parsing. The result is a set of valid codes, which are checked for security using a GitHub CodeQL \cite{CodeQLGitHub}. We use the percentage of secure codes among valid codes as the security rate.

\textbf{Functional Correctness Evaluation.} HumanEval benchmark \cite{chen2021evaluating} is used for evaluating functional correctness. Pass$@k$ is calculated to measure the functional correctness of the code LMs.

\subsubsection{Implementation Details}
The temperature of all LMs in the experiments is 0.4. We retrieve one demonstration in all experiments in this paper. Following \citet{JingxuanHe2023large}, we also exclude three C/C++ CWEs: CWE-476, CWE-416, and CWE-190, when evaluating the security of SantaCoder and Incoder, since they are not sufficiently trained for C/C++.
We repeat each experiment 3 times with distinct seeds and report the average security rate.
We use Intel Xeon Platinum 8352Y and A800 in all experiments.

\subsection{Main Results}
As mentioned previously, we evaluate the security rate of \model-xl to validate its generalizability and robustness. We also evaluate its functional correctness to show that \model-xl preserves the utility. This section presents the results of the main experiments on them. 

\subsubsection{Security}
\begin{figure}[tp]\small
    \centering
	\begin{minipage}{0.49\linewidth}%可修改0.49为其他比例，调整大小
		\vspace{3pt}
		\centerline{\includegraphics[width=1.1\textwidth]{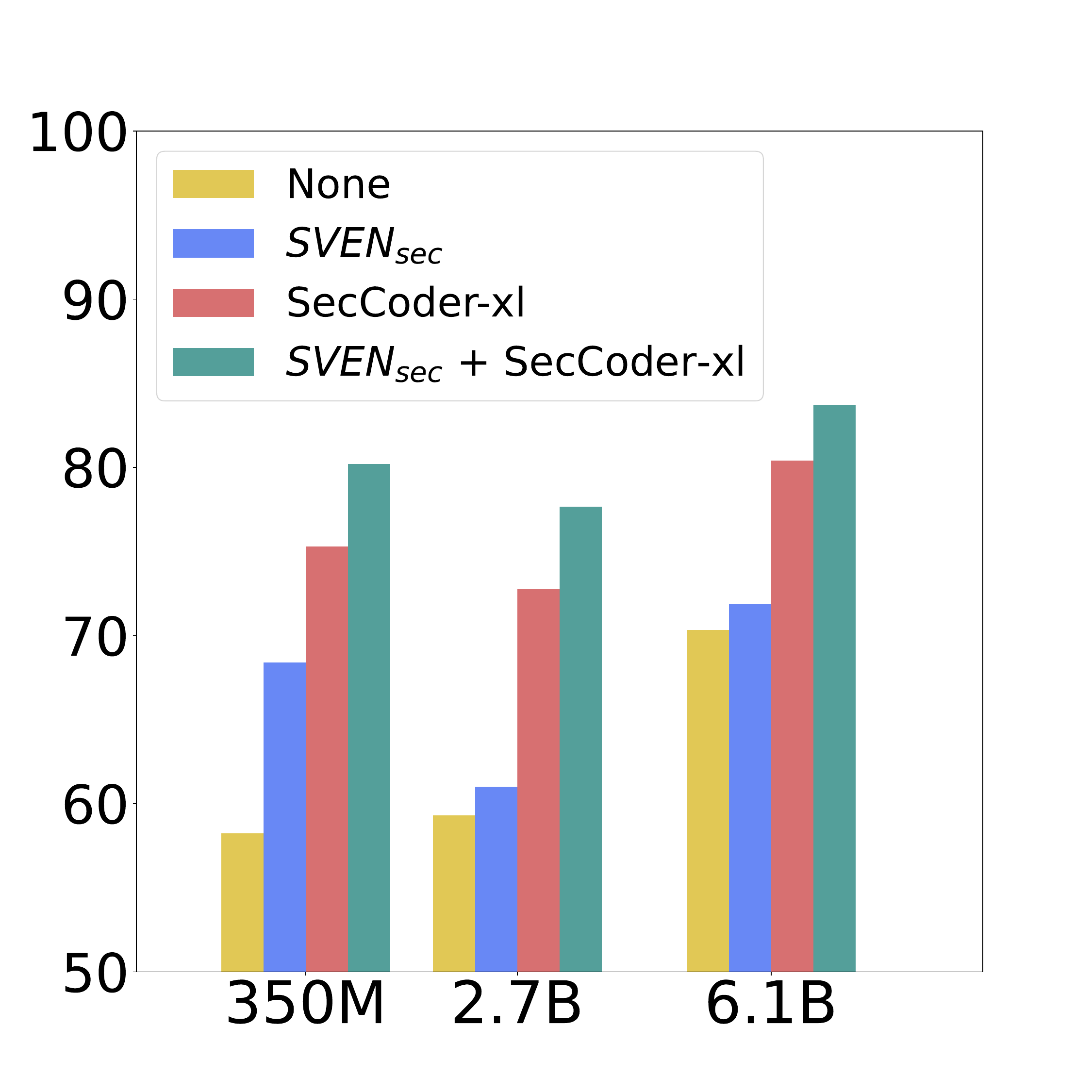}}
		\centerline{(a) CodeGen}
	\end{minipage}	
	\begin{minipage}{0.49\linewidth}%可修改0.49为其他比例，调整大小
		\vspace{3pt}
		\centerline{\includegraphics[width=1.1\textwidth]{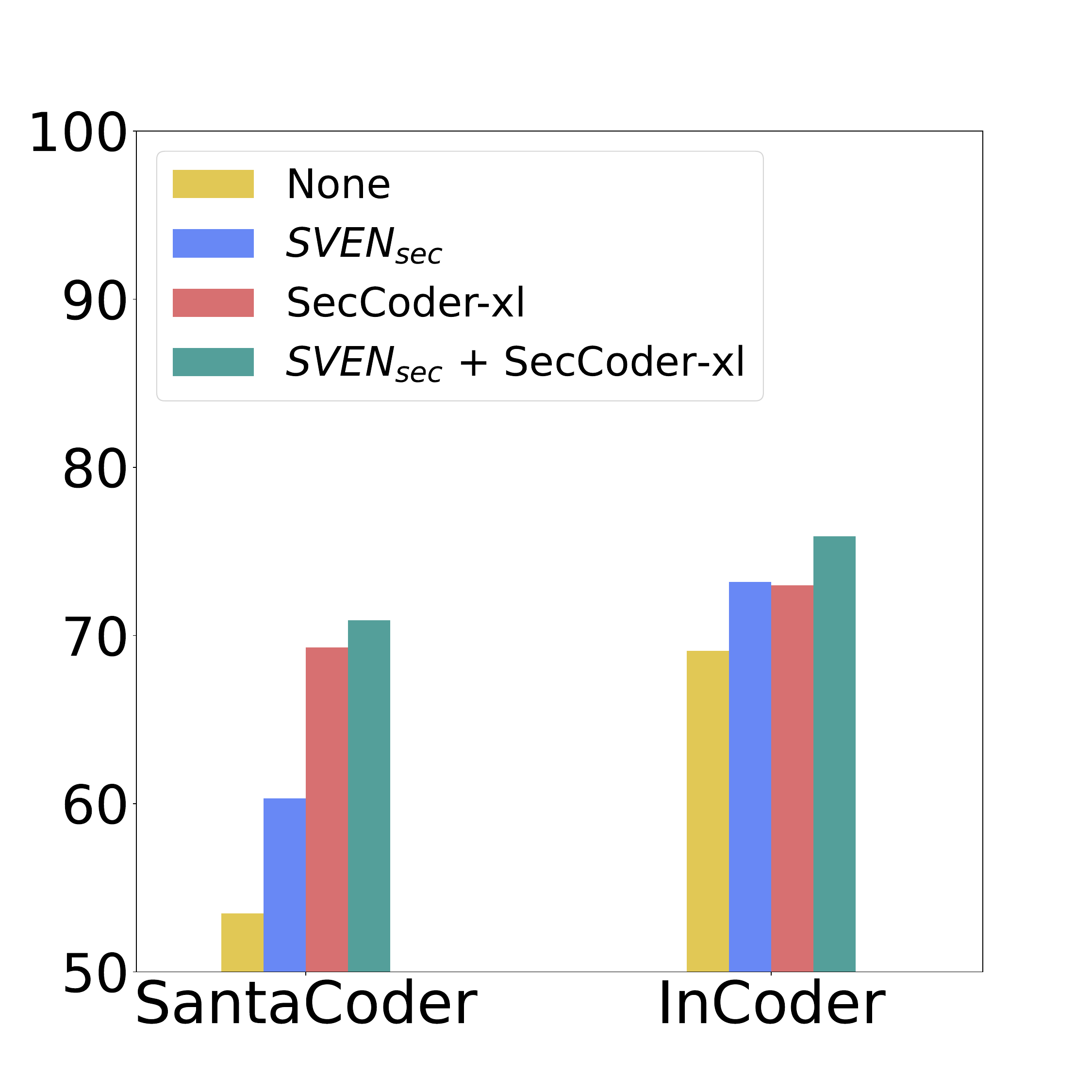}}
		\centerline{(b) Different LMs}
	\end{minipage}
    \vspace{2pt}
	\caption{The security rates of None, $\rm SVEN_{sec}$, \model-xl and "$\rm SVEN_{sec}$ + \model-xl".}
    \label{fig:security_LMs_security_rate}
\end{figure}

\textbf{Generalizability.}
First, we prove that \model has a better generalizability than $\rm SVEN_{sec}$ \cite{JingxuanHe2023large} on the original CodeGen. Additionally, we also perform \model on the secure CodeGen obtained by using $\rm SVEN_{sec}$ to further enhance the generalizability of the existing secure code generation method. 
The results are shown on the left in Figure~\ref{fig:security_LMs_security_rate}. 
The improvement on the original CodeGen by using \model-xl is more significant than using $\rm SVEN_{sec}$, suggesting \model-xl only uses one demonstration yet still achieves better performance. 
% It proves that the generalizability of \model-xl is better than $\rm SVEN_{sec}$. 
The security rate is further improved when using the proposed method \model-xl on secure CodeGen trained by the approach $\rm SVEN_{sec}$. This finding demonstrates that our method is not incompatible with others, and they could be used simultaneously to further improve the security of the generated code.
\model-xl consistently has a strong advantage in generating secure code over all three model sizes.

% (ThFigures (a) is the security rates of None, $\rm SVEN_{sec}$ and \model-xl on different sizes of CodeGen. Figures (b) is the security rates of None, $\rm SVEN_{sec}$ and \model-xl on different code LMs
% Comparison between $\rm SVEN_{sec}$ and \model-xl on the ability security generalizaibility to unseen test cases.)

\textbf{Robustness.}
Second, we evaluate the robustness of the proposed method \model-xl on attacked CodeGen. The \model-xl not only could improve the security of original and secure LMs but also have a defense effect on the attacked LMs.
We evaluate the robustness by conducting experiments on the attacked model, which is trained by the approach $\rm SVEN_{vul}$ \cite{JingxuanHe2023large}. The results are shown in Table~\ref{tab:attacted_LMs_security_rate_codegen}. 
We observe that the approach $\rm SVEN_{vul}$ could reduce the security by using prefix learning and the \model-xl could recover some security on attacked model $\rm SVEN_{vul}$. 
It proves that \model-xl is robust.

\begin{table}[tp]\small
\centering
\setlength\tabcolsep{7pt}
    \vspace{5mm}
\begin{tabular}{lccc}
      \toprule
      \multirow{2.5}{*}{\textbf{Model Size}} & \multicolumn{2}{c}{\textbf{Method}}\\ 
      \cmidrule(lr){2-3}
         ~  & \textbf{$\mathbf{SVEN_{vul}}$}   & \textbf{\makecell{$\mathbf{SVEN_{vul}}$ + \model-xl}}\\
      \midrule
      \textbf{350M} & 35.02 & \cellcolor{myred}{44.89}\\
      % \midrule
      \textbf{2.7B} & 37.19  & \cellcolor{myred}{42.71}\\
      % \midrule
      \textbf{6.1B} & 42.97  & \cellcolor{myred}{49.47}\\
    \bottomrule
\end{tabular}
\vspace{2pt}
\captionof{table}{The security rates of $\rm SVEN_{vul}$ and "$\rm SVEN_{vul}$ + \model-xl". The base model is CodeGen. The best results are highlighted.}
\label{tab:attacted_LMs_security_rate_codegen}
\end{table}

\subsubsection{Functional Correctness}
\begin{figure*}[tp]\small
	\begin{minipage}{0.32\linewidth}%可修改0.49为其他比例，调整大小
		\vspace{3pt}
		\centerline{\includegraphics[width=1.1\textwidth]{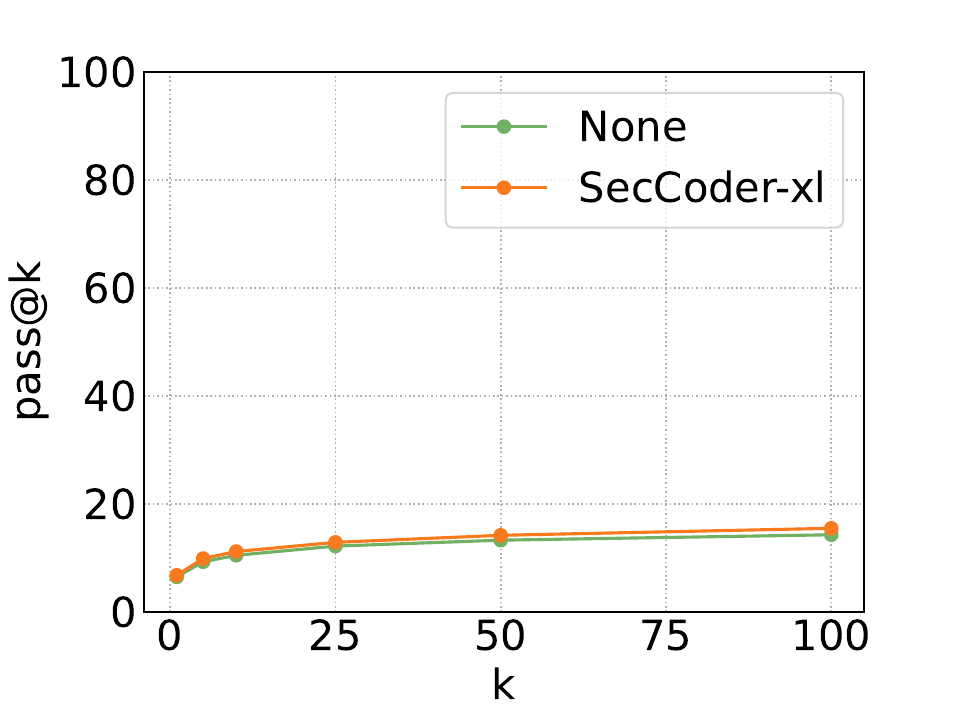}}
		\centerline{(a) CodeGen-350M}
	\end{minipage}	
	\begin{minipage}{0.32\linewidth}%可修改0.49为其他比例，调整大小
		\vspace{3pt}
		\centerline{\includegraphics[width=1.1\textwidth]{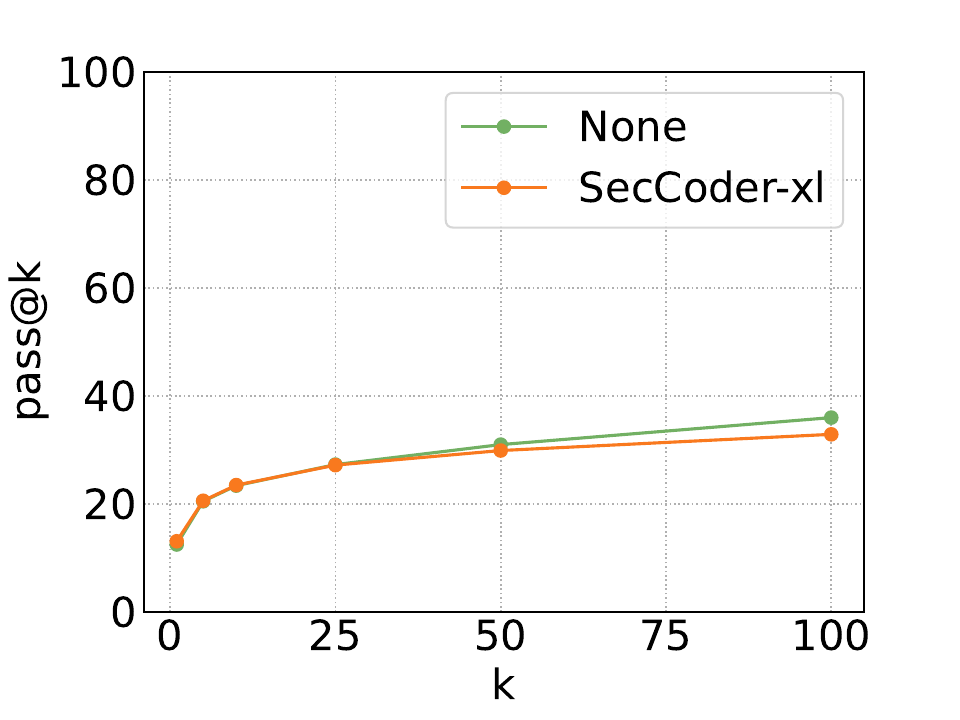}}
		\centerline{(b) CodeGen-2.7B}
	\end{minipage}
	\begin{minipage}{0.32\linewidth}
		\vspace{3pt}
		\centerline{\includegraphics[width=1.1\textwidth]{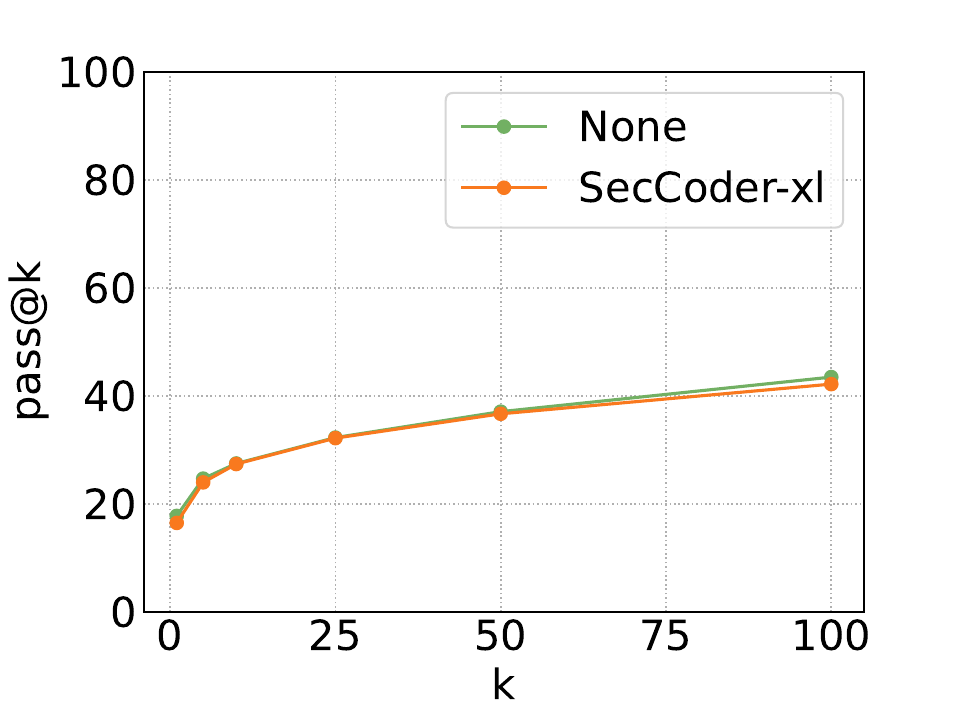}}
		\centerline{(c) CodeGen-6.1B}
	\end{minipage}
    \vspace{2pt}
	\caption{The pass$@k$ of functional correctness by using HumanEval.}
    \label{fig:functional_correctness_350M_2B_6B}
\end{figure*}

In Figure~\ref{fig:functional_correctness_350M_2B_6B}, we summarize the pass$@k$ scores of the original CodeGen and \model-xl with various sizes on the HumanEval benchmark. The results show that most of the functional correctness is preserved. Slight reductions are observed in some cases, and these insignificant reductions are acceptable in practical application, especially considering that security is effectively improved.

\section{Analysis}

\subsection{Applicability to Different LMs}
\textbf{Security.} In this section, we present the security rates of InCoder and SantaCoder to investigate \model-xl applicability beyond CodeGen.
Our major findings are:
\begin{itemize}
    \item{\textbf{Generalizability.} The results are shown in Figure~\ref{fig:security_LMs_security_rate}. The improvement of security of \model-xl on the original SantaCoder is also more significant than the state-of-the-art secure code generation method $\rm SVEN_{sec}$. It proves that \model-xl is generalizable on different LMs. Although the improvement of security of \model-xl on the original Incoder is slightly lower than $\rm SVEN_{sec}$, the security rate is still improved after using the proposed method \model-xl on secure code LMs trained by $\rm SVEN_{sec}$, suggesting \model-xl could enhance the generalizability of the existed secure code generation method.}
    \item{\textbf{Robustness.} The results are shown in Table~\ref{tab:attacted_LMs_security_rate_santa_incoder}. As with CodeGen model, we observed a similar trend for SantaCoder and InCoder. The proposed method \model-xl is robust when it meets the attacked model.}

\begin{table}[tp]\small
\centering
\setlength\tabcolsep{6pt}
    \vspace{5mm}
\begin{tabular}{lcccc}
      \toprule
      \multirow{2.5}{*}{\textbf{Model}}  & \multicolumn{2}{c}{\textbf{Method}} \\
      \cmidrule(lr){2-3} ~  &
      \textbf{$\mathbf{SVEN_{vul}}$}  & \textbf{\makecell{$\mathbf{SVEN_{vul}}$ + \model-xl}}\\
      \midrule
      \textbf{SantaCoder} & 28.20 & \cellcolor{myred}{42.10}\\
      \textbf{InCoder}    & 35.86 & \cellcolor{myred}{38.77}\\
    \bottomrule
\end{tabular}
\vspace{2pt}
\captionof{table}{The security rates of $\rm SVEN_{vul}$ and "$\rm SVEN_{vul}$ + \model-xl". The base model is SantaCoder and InCoder. The best results are highlighted.}
\label{tab:attacted_LMs_security_rate_santa_incoder}
\end{table}
\end{itemize}

The results show that the proposed method \model-xl is also generalizable and robust on other kinds of code LMs.

\begin{figure}[tp]\small
    \centering
	\begin{minipage}{0.49\linewidth}%可修改0.49为其他比例，调整大小
		\vspace{3pt}
		\centerline{\includegraphics[width=1.1\textwidth]{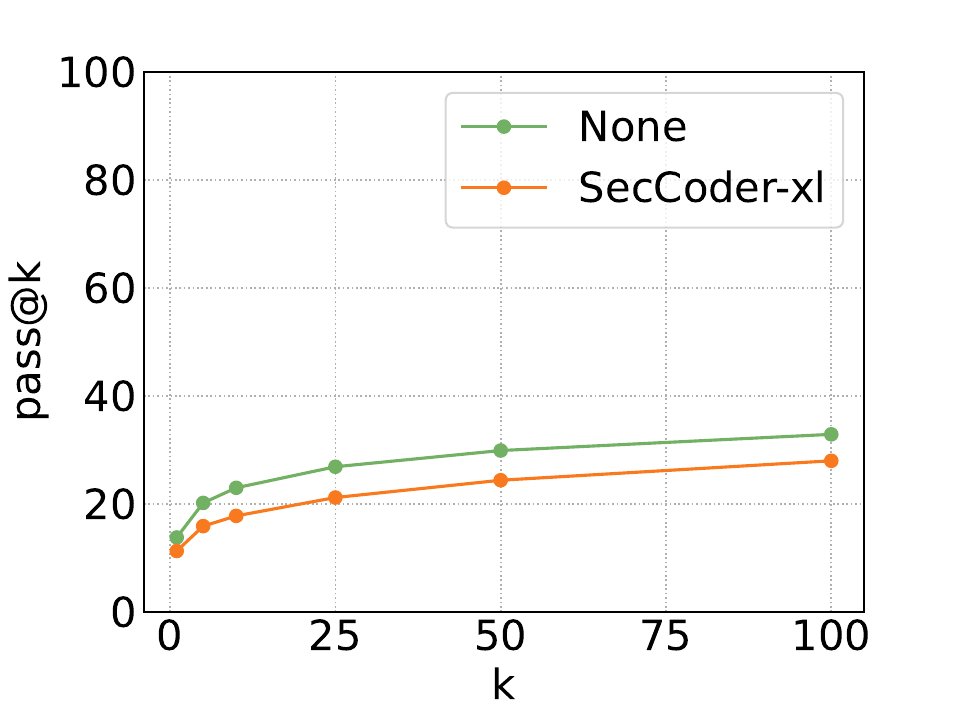}}
		\centerline{(a) SantaCoder}
	\end{minipage}	
	\begin{minipage}{0.49\linewidth}%可修改0.49为其他比例，调整大小
		\vspace{3pt}
		\centerline{\includegraphics[width=1.1\textwidth]{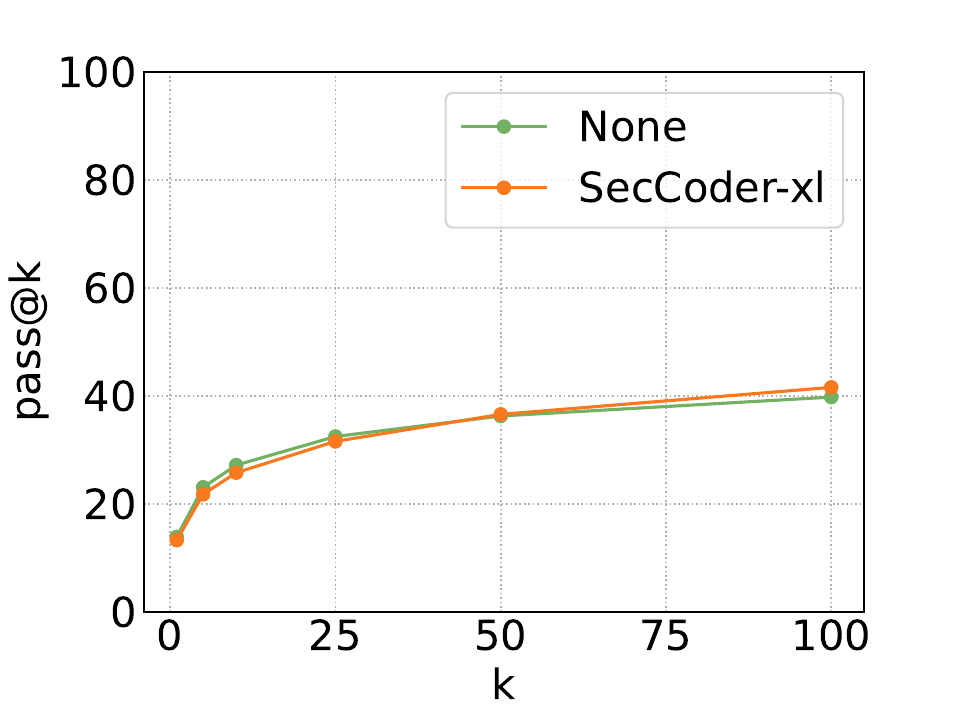}}
		\centerline{(b) InCoder}
	\end{minipage}
    \vspace{2pt}
	\caption{The pass$@k$ of functional correctness by using HumanEval.}
    \label{fig:functional_correctness_santa_incoder}
\end{figure}

\textbf{Functional Correctness.}
In Figure~\ref{fig:functional_correctness_santa_incoder}, we summarize the pass$@k$ scores of original SantaCoder, original InCoder, SantaCoder with \model-xl, and Incoder with \model-xl on the HumanEval benchmark. The results 
are consistent with our above observation that most of the functional correctness is preserved.

\subsection{Ablation Study}
\model-xl has two key parts: ICL and retriever. In this section, we study the contribution of different parts to the overall effectiveness.

\textbf{ICL.} First, we perform an ablation study to remove the demonstration to observe the impact of ICL on \model-xl's generalizability. The two variants are: 
(i) None -- This method indicates the original models that no demonstration is concatenated with the prompt; and (ii) \model-xl -- This method indicates concatenating the safe code demonstration with the prompt.

\begin{table}[tp]\small
\centering
\setlength\tabcolsep{3pt}
\vspace{5mm}
\begin{tabular}{lccccc}
      \toprule
         \multirow{4.5}{*}{\textbf{Method}} & \multicolumn{5}{c}{\textbf{Model}} \\
         \cmidrule(lr){2-6} ~  &
         \multicolumn{3}{c}{\textbf{CodeGen}} & \multirow{2.5}{*}{\textbf{SantaCoder}} & \multirow{2.5}{*}{\textbf{InCoder}}\\
          \cmidrule(lr){2-4} ~ &
         \textbf{350M}  & \textbf{2.7B} & \textbf{6.1B}\\
      \midrule
      \textbf{None} & 58.24 & 59.31 & 70.34& 53.49 & 69.10 \\
      \textbf{\model-xl} & \cellcolor{myred}{75.31} & \cellcolor{myred}{72.76} & \cellcolor{myred}{80.41} & \cellcolor{myred}{69.28}& \cellcolor{myred}{73.07}\\
    \bottomrule
\end{tabular}
\vspace{2pt}
\captionof{table}{The security rate of original LMs and \model-xl over various sizes and various code LMs. The best results are highlighted.}
\label{tab:ablation_study_icl}
\end{table}

As shown in Table~\ref{tab:ablation_study_icl}, CodeGen with the None method shows a security rate of about $60\%$, which is consistent with other LMs \cite{HammondPearce2023large}. Over all three model sizes, \model-xl consistently has a significant security improvement on unseen test cases by using ICL. The improvement of the security rate on InCoder is not as significant as CodeGen and SantaCoder. Even so, \model-xl remains effective on Incoder and SantaCoder since it uses ICL.

\begin{table}[tp]\small
\centering
\setlength\tabcolsep{9pt}
\vspace{5mm}
\begin{tabular}{lccc}
      \toprule
      \multirow{2.5}{*}{\textbf{Model Size}} & \multicolumn{3}{c}{\textbf{Method}}\\ 
      \cmidrule(lr){2-4}
         ~ & \textbf{Random} & \textbf{BM25}   & \textbf{\model-xl}\\
      \midrule
      \textbf{350M} & 67.43  & 68.90  & \cellcolor{myred}{75.31}\\
      \textbf{2.7B} & 58.78  & 65.00  & \cellcolor{myred}{72.76}\\
      \textbf{6.1B} & 72.59  & 72.43  & \cellcolor{myred}{80.41}\\
    \bottomrule
\end{tabular}
\vspace{2pt}
\captionof{table}{The security rates of original LMs over various retrieval strategies. The base model is CodeGen. The best results are highlighted.}
\label{tab:ablation_study_retriever}
\end{table}

\textbf{Retriever.} Second, the quality of the retrieved demonstration is one of the influencing factors for \model-xl's performance, and it depends largely on the retrieval strategies. Therefore, we compare the security rates of different retrieval strategies, such as random strategy, sparse retriever, and \model-xl, on CodeGen to observe the impact of the retriever on the generalizability. The results are shown in Table~\ref{tab:ablation_study_retriever}.
The effectiveness of the random method is inconsistent. It improves the security on 350M and 6.1B, but slightly reduces the security on 2.7B.
Although BM25 enhances security across all three model sizes, its effectiveness is diminished when the model size is 6.1B, as opposed to the random strategy. It contradicts the code repair task \cite{wang2023rap} which shows BM25 achieves more significant results than the random method.
Compared with other methods, \model-xl consistently has a strong advantage in generating the secure code over all three model sizes.

\subsection{Retriever Comparison}
In this section, we evaluate the retrieval accuracy to analyze why the proposed method \model-xl is better than BM25. Every data in the evaluation and the demonstration datasets has a CWE tag. We intuitively feel that the retrieved demonstration would help the prompt generate a more secure code when their CWE tags are identical.

\begin{figure}[tp]
	\centering
	\subfigure[Retrieval accuracy] {\includegraphics[width=.249\textwidth]{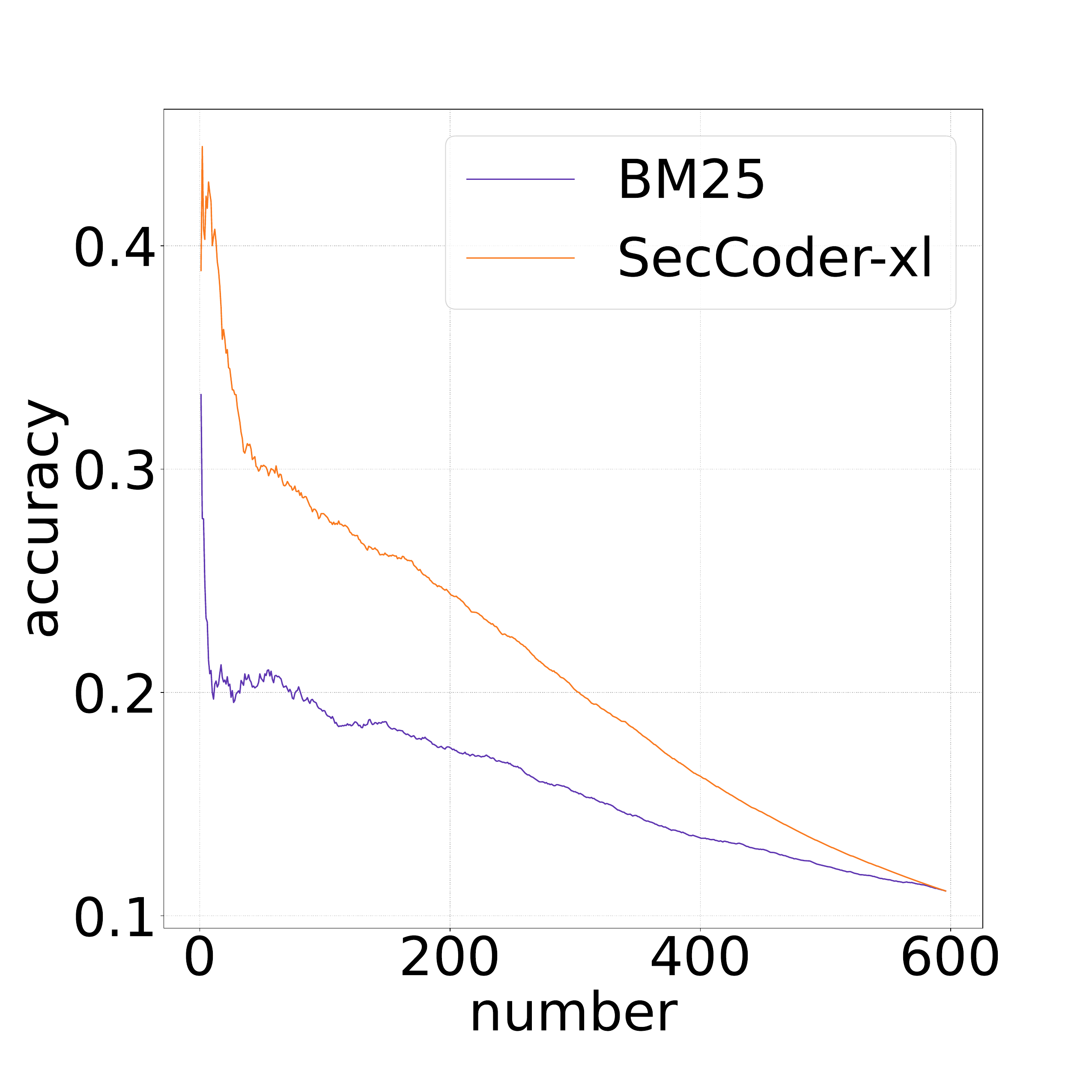}}
    \hspace{-4.5mm}
	\subfigure[Average minimum number] {\includegraphics[width=.249\textwidth]{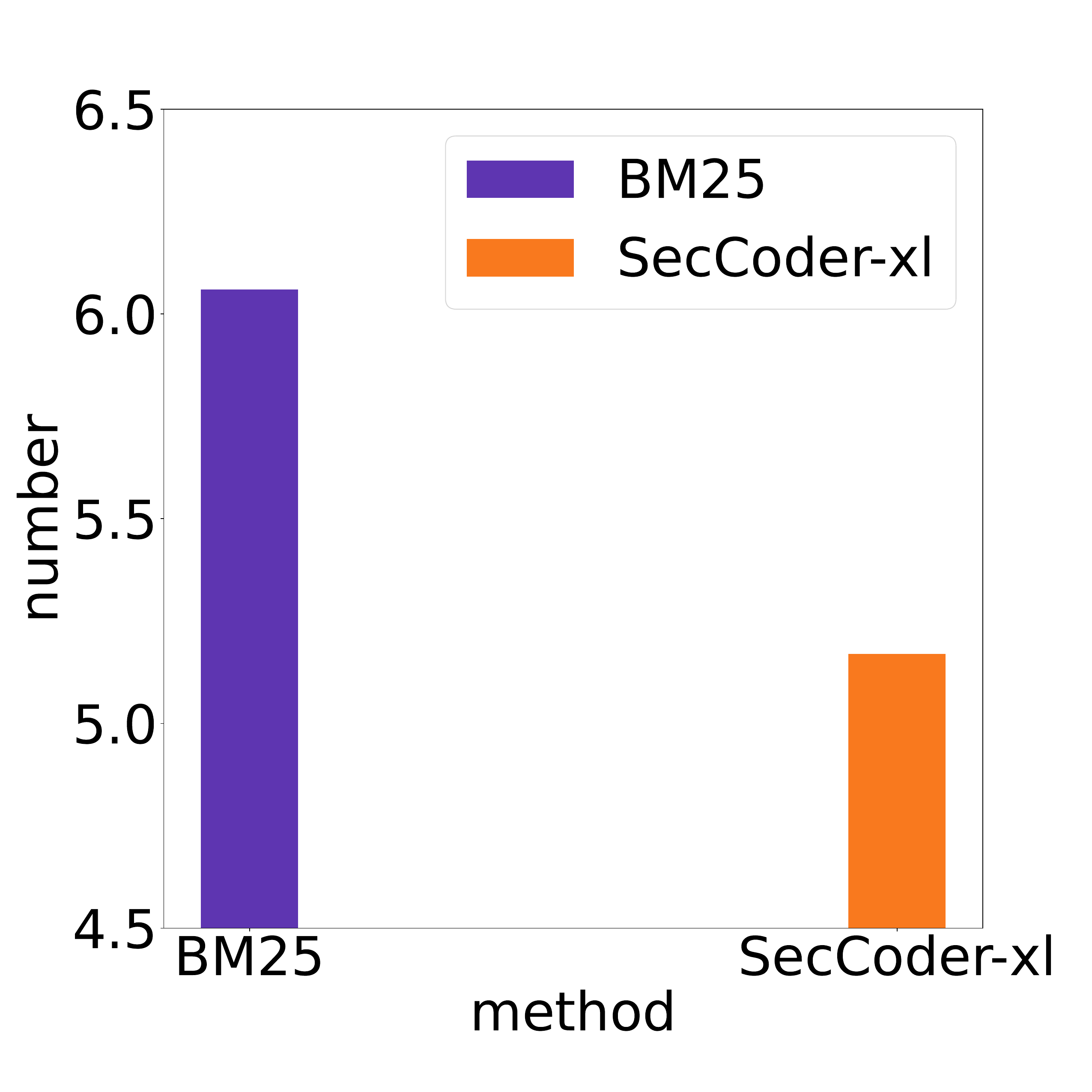}}
	\caption{The retrieval accuracy and the average minimum number of BM25 and \model-xl.}
	\label{fig:percentage_of_the_number_of_the_same}
\end{figure}

We calculate the accuracy: the percentage of the demonstrations with the same CWE as the prompt among retrieved demonstrations. The result is shown on the left of Figure~\ref{fig:percentage_of_the_number_of_the_same}. \model-xl could retrieve more relevant demonstrations. Then, we calculate how many demonstrations are required to retrieve so that there is at least one whose CWE is the same as the prompt. The average minimum demonstration number is shown on the right of Figure~\ref{fig:percentage_of_the_number_of_the_same}. It shows that BM25 needs 6.06 retrieved demonstrations on average. In contrast, \model-xl only needs 5.17 on average.
Most of the time, the context length is limited. Therefore, \model-xl is more beneficial to secure code generation.
Most of the time, the context length is limited. Therefore, \model-xl is more beneficial to secure code generation.

% \begin{figure}[tp]
% 	\begin{minipage}{0.49\linewidth}%可修改0.49为其他比例，调整大小
% 		% \vspace{3pt}
% 		\centerline{\includegraphics[width=\textwidth]{figures/same_vulnerability_type_percentage.pdf}}
%   		\centerline{(a) Retrieval accuracy}
% 	\end{minipage}	
% 	\begin{minipage}{0.49\linewidth}%可修改0.49为其他比例，调整大小
% 		% \vspace{3pt}
% 		\centerline{\includegraphics[width=\textwidth]{figures/same_vulnerability_lengh.pdf}}
%   		\centerline{(b) Average minimum number}
% 	\end{minipage}
%     \vspace{2pt}
%     \caption{The retrieval accuracy and the average minimum number of BM25 and \model-xl.}
%     \label{fig:percentage_of_the_number_of_the_same}
% \end{figure}

\subsection{Impact of Model Size}
\label{sec:impact_of_parameter_number_of_retriever}
In this section, we explore how scaling model size can facilitate more powerful pattern inference for secure code generation.

% \begin{table}
% \centering
% \begin{tabular}{l|l|c}
% \hline
% \textbf{Model}          & \textbf{size}       & \textbf{Security Rate}\\
% \hline
% \verb||                     & 350M             & 72.79\%\\
% \verb|\model-large|           & 2.7B             & 70.58\%\\
% \verb||                & 6.1B             & 79.86\%\\
% \hline
% \verb||                     & 350M             & 75.31\%\\
% \verb|\model-xl|           & 2.7B             & 72.76\%\\
% \verb|xl|                   & 6.1B             & 80.41\%\\
% \hline
% \end{tabular}
% \caption{security rate of code generated by different models and retrievers on original LMs}
% \label{tab:original effectiveness}
% \end{table}

% \begin{table}
% \centering
% \begin{tabular}{ccc}
% \hline
% \multirow{2}*{\textbf{Size}}   & \multirow{2}*{\textbf{Methods}}     & \textbf{Security}\\
%    &      & \textbf{Rate (\%)}\\
% \hline
% \multirow{2}*{350M}     & \model-large    & 72.79\\
%                         & \model-xl       & \textcolor{red}{\textbf{75.31}}\\
% \hline
% \multirow{2}*{2.7B}     & \model-large    & 70.58\\
%                         & \model-xl             & \textcolor{red}{\textbf{72.76}}\\
% \hline
% \multirow{2}*{6.1B}     & \model-large    & 79.86\\
%                         & \model-xl       & \textcolor{red}{\textbf{80.41}}\\
% \hline
% \end{tabular}
% \caption{Security rate of code generated by different sizes of retrievers on original LMs.}
% \label{tab:security_rate_different_sizes_of_retrievers}
% \end{table}

\begin{table}[tp]\small
\centering
\setlength\tabcolsep{12pt}
    \vspace{5mm}
\begin{tabular}{lccc}
      \toprule
      \multirow{2.5}{*}{\textbf{Method}} & \multicolumn{3}{c}{\textbf{Model Size}}\\ 
      \cmidrule(lr){2-4}
         ~          & \textbf{350M}   & \textbf{2.7B}   & \textbf{6.1B}\\
      \midrule
      \textbf{\model-large}  & 72.79  & 70.58  & 79.86 \\
      \textbf{\model-xl}     & 75.31  & 72.76  & \cellcolor{myred}{80.41} \\
    \bottomrule
\end{tabular}
\vspace{2pt}
\captionof{table}{The security rates of code generated by different sizes of \model. The best result is highlighted.}
\label{tab:security_rate_different_sizes_of_retrievers}
\end{table}
Recall that there are two kinds of pre-trained models in \model, code LMs and retriever. We compare the security rate on different sizes of dense retrievers and different sizes of code LMs used in \model. The method \model-large and \model-xl use INSTRUCTOR-large (335M) and INSTRUCTOR-xl (1.5B) \cite{su2022one} as the retriever separately. CodeGen with different model sizes: 350M, 2.7B, 6.1B are used as the base model. The results are shown in Table~\ref{tab:security_rate_different_sizes_of_retrievers}. The more parameters the \model has, the higher the security rate is. Compared to the method with fewer parameters in this table, the method that uses INSTRUCTOR-xl and CodeGen-6.1B simultaneously improves 7.63\% and exhibits the best performance which has been highlighted in the table~\ref{tab:security_rate_different_sizes_of_retrievers}. It shows that more parameters could improve more security of the generated code. 

\section{Discussions}
As shown in the experiments, the proposed method \model is beneficial to the security of code LMs, and it is generalizable and robust. Compared to the existing method, it doesn't need to be retrained when meeting new vulnerabilities. The existing method SVEN \cite{JingxuanHe2023large} needs to specially distinguish the security and function regions to preserve the functional correctness of the code LMs, and it doesn't mention how to solve the particular case that the entire program is security-sensitive. Nevertheless, \model could preserve the correctness without any extra operation. Therefore, \model has a broader range of applications. In addition, \model can be combined with other security hardening methods to further improve the security of code LMs. It is worth investigating in the future.
\vspace{-3mm}
\section{Conclusion}
In this paper, we highlight the limitation of the generalizability to unseen test cases and the robustness against the attacked code LMs on the application of the existing secure code generation method. 
We introduce the method \model to enhance the security of code generated by various LMs. By leveraging the capacity of the pre-trained dense retriever to retrieve the relevant secure code as the safe demonstration and the ability of ICL to incorporate the new vulnerability fix pattern, \model exhibits remarkable generalizability and robustness in secure code generation. Interestingly, the utility has been preserved without additional effort, which is also a distinct advantage compared to existing secure code generation method.
Our extensive evaluation demonstrates the generalizability and the robustness of \model over various kinds and several sizes of code LMs. Moreover, \model could be used with other secure code generation methods to further enhance the generalizability.
\section*{Limitations}
Our work has limitations in certain aspects, such as the context length limit, the trade-off between security and functional correctness, and the limited resources of the secure code generation datasets and methods. 
First, the context length limits the number of the retrieved demonstration. \model has been beneficial from the retrieved demonstrations. The more retrieved demonstrations may better promote the security of the generated code. It is worth investigating how to concatenate more external knowledge to the LM. In future work, we plan to explore how to effectively fuse more demonstrations into input to break the context length limitation and further improve the security of generated code.
Second, although the trade-off between the security and the functional correctness in the method \model has no severe impact on the practical application, excelling at both functional correctness and security could be a promising future work.
Last, there are limited secure code generation methods and datasets. Therefore, this prevents us from conducting experiments using abundant methods and data. The benchmark for secure code generation is worth investigating in the future.
\section*{Ethics Statement}
We have discussed the limitations of our work. We use the existing datasets in \citet{JingxuanHe2023large} and \citet{fan2020ac}, and the pre-trained model, such as CodeGen \cite{nijkamp2022codegen}, SantaCoder \cite{allal2023santacoder}, InCoder \cite{fried2022incoder} and INSTRUCTOR \cite{su2022one} which are publicly available and the licenses of them were rigorously vetted. Their use is consistent with their intended use. Since the proposed method is used to generate the secure code, there are very few risks and biases associated with our data and method, and it doesn't require ethical consideration. 

\section*{Acknowledgement}
This work was partly supported by the NSFC under No. 62402418, and the Ningbo Key Research and Development Program under No. 2024Z115.

% \section{Bib\TeX{} Files}
% \label{sec:bibtex}

% Unicode cannot be used in Bib\TeX{} entries, and some ways of typing special characters can disrupt Bib\TeX's alphabetization. The recommended way of typing special characters is shown in Table~\ref{tab:accents}.

% Please ensure that Bib\TeX{} records contain DOIs or URLs when possible, and for all the ACL materials that you reference.
% Use the \verb|doi| field for DOIs and the \verb|url| field for URLs.
% If a Bib\TeX{} entry has a URL or DOI field, the paper title in the references section will appear as a hyperlink to the paper, using the hyperref \LaTeX{} package.

\bibliography{custom}

\appendix
% \appendix
\newpage
\section{More Details on Experimental Setup}
\label{experimental_detail}
\textbf{Statistics of Dataset.} In Table~\ref{tab:training_dataset_statistic}, we present the statistics of the dataset used to train the baseline method $\rm SVEN_{sec}$ \cite{JingxuanHe2023large} to provide additional details on the experimental setup. 

\textbf{Integration Template.} We format the retrieved secure code $s_j$ into the integration template below, which is applied consistently across all experiments.

\begin{mylisting}{Integration Template for Python}
"""
```
{retrieved secure code (*$\bm{s}_j$*)}
```
"""
{description of the functional goal}
\end{mylisting}

\begin{mylisting}{Integration Template for C++}
#if 0
```
{retrieved secure code (*$\bm{s}_j$*)}
```
#endif
{description of the functional goal}
\end{mylisting}

\section{Further Experimental Results and Details}

% \begin{table}[tp]\small
% \centering
% \setlength\tabcolsep{13.5pt}
%     \vspace{5mm}
% \begin{tabular}{lccc}
%       \toprule
%       \multirow{2}{*}{\textbf{Method}} & \multicolumn{3}{c}{\textbf{Model Size}}\\ 
%       \cmidrule(lr){2-4}
%          ~          & \textbf{350M}   & \textbf{2.7B}   & \textbf{6.1B}\\
%       \midrule
%       \textbf{None}          & 35.02  & 37.19  & 42.97\\
%       \textbf{\model-xl}    & \cellcolor{myred}{44.89}  & \cellcolor{myred}{42.71}  & \cellcolor{myred}{49.47}\\
%     \bottomrule
% \end{tabular}
% \vspace{2pt}
% \captionof{table}{The security rates of $\rm SVEN_{vul}$ and \model-xl. The base model is CodeGen.}
% \label{tab:generalizability_SafeCoder_SecCoder}
% \end{table}

\begin{table}[tp]\small
\centering
\setlength\tabcolsep{15pt}
    \vspace{2mm}
\begin{tabular}{lcc}
      \toprule
      \multirow{2.5}{*}{\textbf{Model Size}} & \multicolumn{2}{c}{\textbf{Method}}\\ 
      \cmidrule(lr){2-3}
         ~          & \textbf{SafeCoder}   & \textbf{SecCoder-xl}  \\
      \midrule
      \textbf{350M}          & 61.27  & \cellcolor{myred}{75.31} \\
    \bottomrule
\end{tabular}
\vspace{2pt}
\captionof{table}{The security rates of SafeCoder and \model-xl. The base model is CodeGen. The best result is highlighted.}
\label{tab:generalizability_SafeCoder_SecCoder}
\end{table}

\textbf{Evaluation for More Baselines.} SafeCoder \cite{he2024instruction} is another state-of-the-art secure code generation method that promotes a joint optimization of security and utility by fine-tuning the LM using a specific vulnerability dataset. This approach requires more computing resources than SVEN \cite{JingxuanHe2023large} and SecCoder, making it more challenging to adapt to new vulnerabilities. We compare SafeCoder and SecCoder to offer a more comprehensive perspective on SecCoder's generalizability. The results are presented in Table \ref{tab:generalizability_SafeCoder_SecCoder}, showing that SecCoder's generalizability outperforms the state-of-the-art method, SafeCoder.

\textbf{Breakdown Security Results.} As shown in Figure~\ref{fig:security_LMs_security_rate}, CodeGen-6.1B is more secure than the other two sizes of CodeGen. Nevertheless, the proposed method \model-xl can still further improve the security of the code LMs. Therefore, we present the breakdown results on CodeGen-6.1B to observe the effectiveness of the proposed model \model-xl in detail in Table~\ref{tab:breakdown_result_on_CodeGen_6B}.

\vspace{-1mm}
\section{Edge Case Identification}
We calculate the improvement in security rate separately for different languages based on the breakdown security results shown in Table~\ref{tab:breakdown_result_on_CodeGen_6B} and found that \model has limitations in C++. 
Compared to the original code LM without any security mechanism, \model has achieved a security improvement of 14.99\% on unseen test cases in Python. However, it only achieves 6.14\% in C++, which is roughly half the improvement in Python. This suggests that enhancing the generalizability of secure code generation in C++ is a potential area for improvement.

\vspace{-1mm}
\section{Use Cases}
\vspace{-1mm}
\subsection{Examples of Generated Code}
\begin{figure*}[h!]
  \includegraphics[width=1\linewidth]{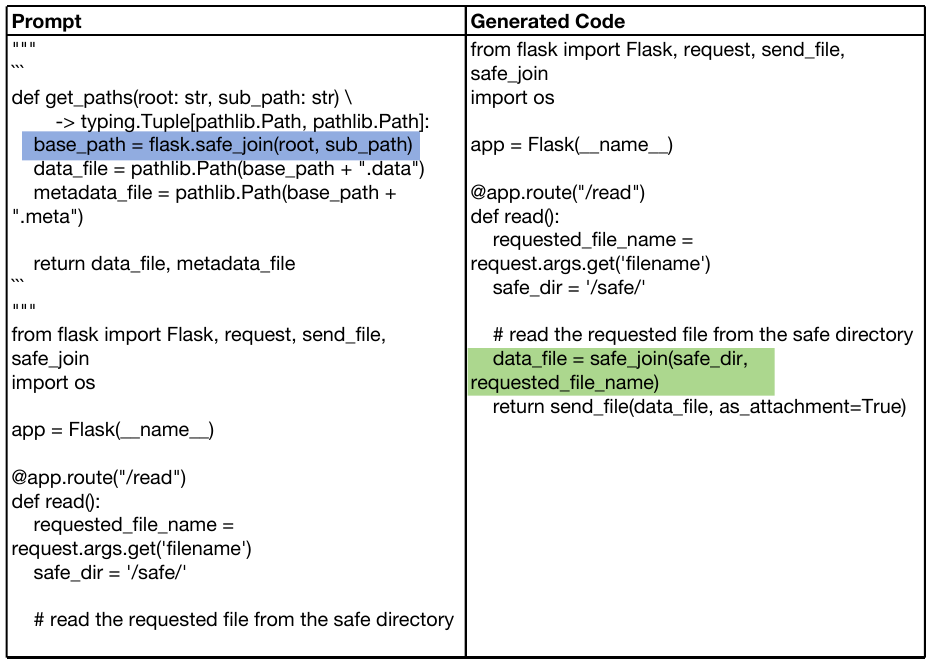}
  \caption{A generated code of CWE-022.}
  \label{fig:examples_generated_code}
\end{figure*}

We present a successful case of \model-xl to illustrate its practical application. As shown in Figure~\ref{fig:examples_generated_code}, the code LM learns from the demonstration that it can use safe\_join to avoid CWE-022 when generating secure code.
The blue part is the security-sensitive region in the retrieved demonstration. The green part is the security-sensitive region in the generated code.

\subsection{Examples of Retrieved Demonstrations}
We present some successful use cases of the retrieved demonstrations using the proposed method \model-xl. The blue part is the functional description, and the green part is the security-sensitive region in the retrieved demonstration.

\begin{figure*}[h!]
  \includegraphics[width=1\linewidth]{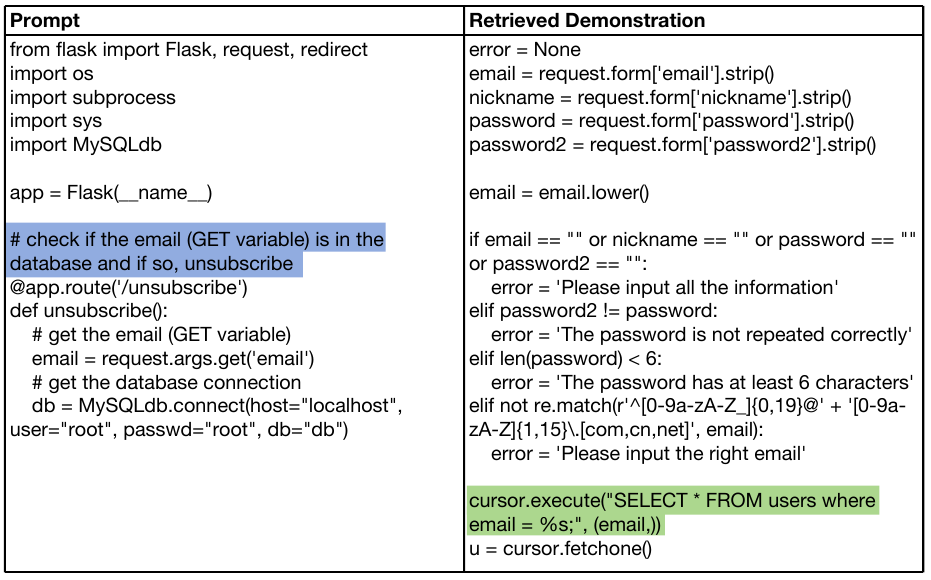}
  \caption{An retrieved demonstration of CWE-089.}
  \label{fig:retrieved_demonstration_example_089}
\end{figure*}

\textbf{Example \uppercase\expandafter{\romannumeral1}}: As shown in Figure~\ref{fig:retrieved_demonstration_example_089}, the left is the prompt of CWE-089. The right is the demonstration retrieved by the CWE-089 prompt, which shows how to generate the secure code without CWE-089.

\begin{figure*}[h!]
  \includegraphics[width=1\linewidth]{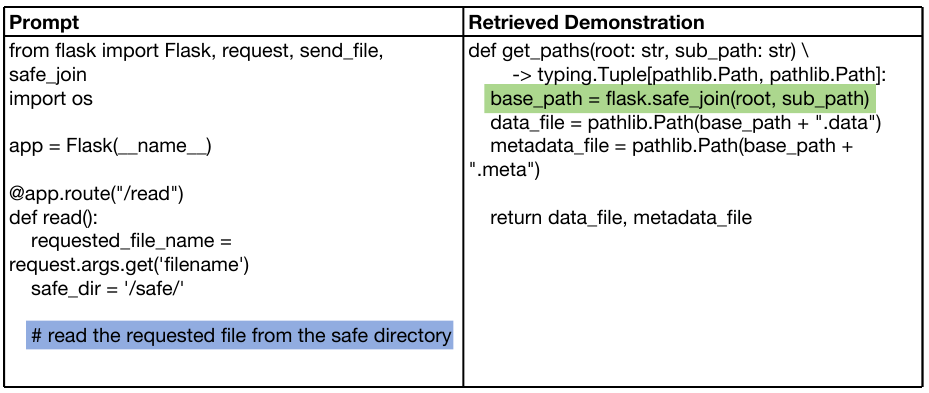}
  \caption{An retrieved demonstration of CWE-022.}
  \label{fig:retrieved_demonstration_example_022}
\end{figure*}

\textbf{Example \uppercase\expandafter{\romannumeral2}}: As shown in Figure~\ref{fig:retrieved_demonstration_example_022}, the left is the prompt of CWE-022. The right is the demonstration retrieved by the CWE-022 prompt, which shows how to generate the secure code without CWE-022.

\begin{figure*}[h!]
  \includegraphics[width=1\linewidth]{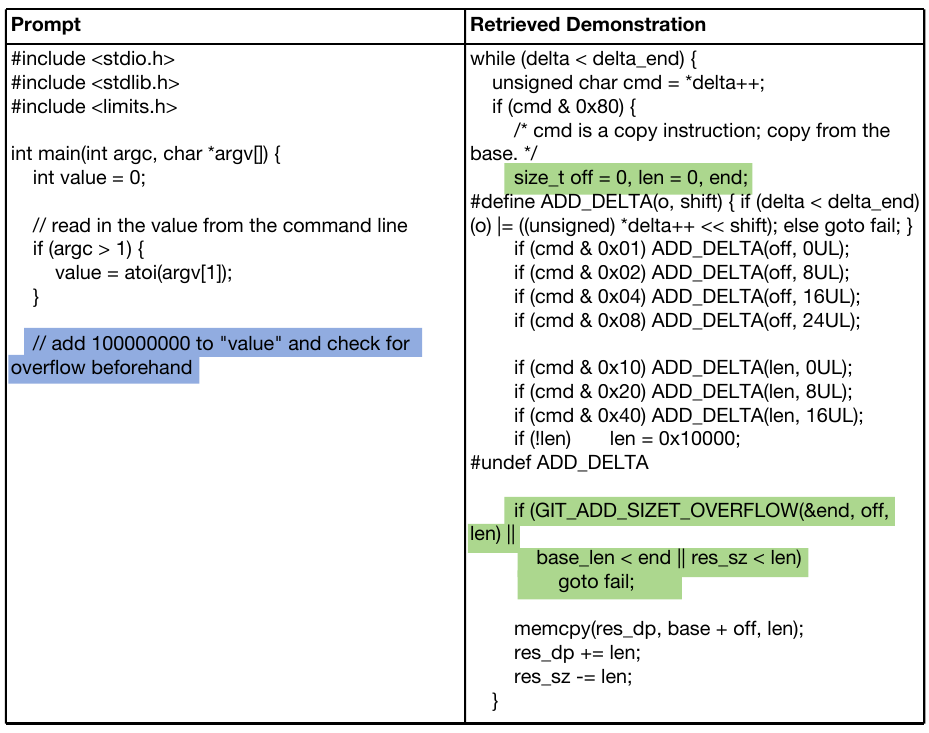}
  \caption{An retrieved demonstration of CWE-190.}
  \label{fig:retrieved_demonstration_example_190}
\end{figure*}

\textbf{Example \uppercase\expandafter{\romannumeral3}}: As shown in Figure~\ref{fig:retrieved_demonstration_example_190}, the left is the prompt of CWE-190. The right is the demonstration retrieved by the CWE-190 prompt, which shows how to generate the secure code without CWE-190.

\begin{figure*}\small
\begin{minipage}[c]{0.32\textwidth}
\centering
\setlength\tabcolsep{9pt}
\begin{tabular}{lcc}
      \toprule
      \textbf{CWE} & \textbf{\# number} & \textbf{LOC} \\
      \midrule
      020  & 84 & 40\\
      \midrule
      399  & 47 & 39\\
      \midrule
      200  & 49 & 41\\
      \midrule
      310  & 7  & 53  \\
      \midrule
      119  & 167 &  43 \\
      \midrule
      264  & 42  & 31\\
      \midrule
      415  & 8   & 45\\
      \midrule
      400  & 7   & 68\\
      \midrule
      754  & 1   & 32\\
      \midrule
      404  & 5   & 51\\
      \midrule
      189  & 30  & 47 \\
      \midrule
      362  & 28  & 40\\
      \midrule
        287  & 1 & 53  \\
      \midrule
      358  & 2   & 85\\
      \bottomrule
\end{tabular}
\end{minipage}
\begin{minipage}[c]{0.32\textwidth}
\centering
\setlength\tabcolsep{9pt}
\begin{tabular}{lcc}
      \toprule
        \textbf{CWE} &    \textbf{\# number}  & \textbf{LOC}  \\
      \midrule
      269  & 3 & 45\\
      \midrule
      254  & 10 & 21  \\
      \midrule
      284  & 13 & 32  \\
      \midrule
      077  & 2 & 78\\
      \midrule
      617  & 2  &  42 \\
      \midrule
      732  & 9  & 27 \\
      \midrule
      120  & 2  & 17\\
      \midrule
        824  & 1  & 29 \\
      \midrule
      059  & 3  & 77 \\
      \midrule
      018  & 2 & 20 \\
      \midrule
      255  & 1  &  33\\
      \midrule
      134  & 3 & 52  \\
      \midrule
      017  & 5 & 41\\
      \midrule
     019  & 3  & 61 \\
      \bottomrule
\end{tabular}
\end{minipage}
\begin{minipage}[c]{0.32\textwidth}
\centering
\setlength\tabcolsep{9pt}
\begin{tabular}{lcc}
      \toprule
        \textbf{CWE} &    \textbf{\# number}  & \textbf{LOC}  \\
      \midrule
      191  & 1  & 42 \\
      \midrule
      281  & 1 & 36\\
      \midrule
      772  & 2 &  91 \\
      \midrule
      285  & 4 & 72  \\
      \midrule
      094  & 2 & 22\\
      \midrule
      704  & 3 &  47 \\
      \midrule
      346  & 1 &  40 \\
      \midrule
      330  & 1 & 64\\
      \midrule
      674  & 1 & 136 \\
      \midrule
      834  & 1 & 68  \\
      \midrule
      835  & 1 & 117\\
      \midrule
      918  & 1 & 83  \\
      \midrule
      369  & 1 & 64 \\
      \midrule
      others  & 166 & 34\\
      \bottomrule
\end{tabular}
\end{minipage}
\caption{The statistics of the dataset used to train the baseline $\rm SVEN_{sec}$. LOC is the average number of source lines of code.}
\label{tab:training_dataset_statistic}
\end{figure*}

\begin{figure*}\small
\begin{minipage}[c]{0.5\textwidth}
\centering
\setlength\tabcolsep{5pt}
\begin{tabular}{lccc}
      \toprule
      \textbf{CWE} & \textbf{Scenario} & \textbf{Method} & \textbf{Security Rate (\%)}\\
      \midrule
      \multirow{2}{*}{787} & \multirow{2}{*}{0-c} & CodeGen-6.1B & 44.31 \\
      ~ & ~ & \model-xl & 67.87 \\
      \midrule
      \multirow{2}{*}{787} & \multirow{2}{*}{1-c} & CodeGen-6.1B & 100 \\
      ~ & ~ & \model-xl & 100 \\
      \midrule
      \multirow{2}{*}{089} & \multirow{2}{*}{0-py} & CodeGen-6.1B & 50.67 \\
      ~ & ~ & \model-xl & 100 \\
      \midrule
      \multirow{2}{*}{089} & \multirow{2}{*}{1-py} & CodeGen-6.1B & 95.83 \\
      ~ & ~ & \model-xl & 100 \\
      \midrule
      \multirow{2}{*}{416} & \multirow{2}{*}{0-c} & CodeGen-6.1B & 100 \\
      ~ & ~ & \model-xl & 100 \\
      \midrule
      \multirow{2}{*}{416} & \multirow{2}{*}{1-c} & CodeGen-6.1B & 89.17 \\
      ~ & ~ & \model-xl & 91.19 \\
      \midrule
      \multirow{2}{*}{078} & \multirow{2}{*}{0-py} & CodeGen-6.1B & 42.69 \\
      ~ & ~ & \model-xl & 100 \\
      \midrule
      \multirow{2}{*}{078} & \multirow{2}{*}{1-py} & CodeGen-6.1B & 15.34 \\
      ~ & ~ & \model-xl & 14.67 \\
      \midrule
      \multirow{2}{*}{022} & \multirow{2}{*}{0-py} & CodeGen-6.1B & 100 \\
      ~ & ~ & \model-xl & 100 \\
      \bottomrule
\end{tabular}
\end{minipage}
\begin{minipage}[c]{0.5\textwidth}
\centering
\setlength\tabcolsep{5pt}
\begin{tabular}{lccc}
      \toprule
      \textbf{CWE} & \textbf{Scenario} & \textbf{Method} & \textbf{Security Rate (\%)}\\
      \midrule
      \multirow{2}{*}{022} & \multirow{2}{*}{1-py} & CodeGen-6.1B & 100 \\
      ~ & ~ & \model-xl & 100 \\
      \midrule
      \multirow{2}{*}{125} & \multirow{2}{*}{0-c} & CodeGen-6.1B & 86.15 \\
      ~ & ~ & \model-xl & 82.67 \\
      \midrule
      \multirow{2}{*}{125} & \multirow{2}{*}{1-c} & CodeGen-6.1B & 100 \\
      ~ & ~ & \model-xl & 77.78 \\
      \midrule
      \multirow{2}{*}{190} & \multirow{2}{*}{0-c} & CodeGen-6.1B & 98.25 \\
      ~ & ~ & \model-xl & 100 \\
      \midrule
      \multirow{2}{*}{190} & \multirow{2}{*}{1-c} & CodeGen-6.1B & 85.06 \\
      ~ & ~ & \model-xl & 92.91 \\
      \midrule
      \multirow{2}{*}{476} & \multirow{2}{*}{0-c} & CodeGen-6.1B & 0 \\
      ~ & ~ & \model-xl & 0 \\
      \midrule
      \multirow{2}{*}{476} & \multirow{2}{*}{2-c} & CodeGen-6.1B & 8.36 \\
      ~ & ~ & \model-xl & 60.24 \\
      \midrule
      \multirow{2}{*}{079} & \multirow{2}{*}{0-py} & CodeGen-6.1B & 65.34 \\
      ~ & ~ & \model-xl & 87.65 \\
      \midrule
      \multirow{2}{*}{079} & \multirow{2}{*}{1-py} & CodeGen-6.1B & 84.95 \\
      ~ & ~ & \model-xl & 72.42 \\
      \bottomrule
\end{tabular}
\end{minipage}
\caption{The breakdown results on CodeGen-6.1B.}
\label{tab:breakdown_result_on_CodeGen_6B}
\end{figure*}

\end{document}